\documentclass[prl,aps,twocolumn]{revtex4}
\usepackage{amsmath}
\usepackage{amssymb}
\usepackage {graphicx}
\usepackage{color}

\newcommand{\beq}{\begin{equation}}
\newcommand{\eeq}{\end{equation}}
\newcommand{\bea}{\begin{eqnarray}}
\newcommand{\eea}{\end{eqnarray}}
\newcommand{\e}{\varepsilon}

\newcommand{\bk}{{\vec k}}

\newcommand{\nn}{\nonumber}
\newcommand{\bse}{\begin{subequations}}
\newcommand{\ese}{\end{subequations}}
\newcommand{\bwt}{\begin{widetext}}
\newcommand{\ewt}{\end{widetext}}

\newcommand{\bsu}{\begin{subequations}}
\newcommand{\esu}{\end{subequations}}

\begin{document}
\title{Electron Spin Resonance\\ in a Two-Dimensional Fermi Liquid with Spin-Orbit Coupling}
\author{Saurabh Maiti$,^{1,2}$ Muhammad Imran,$^1$ and Dmitrii L. Maslov$^1$}
\affiliation {~$^1$Department of Physics, University of Florida, Gainesville, FL 32611}
\affiliation {~$^2$National High Magnetic Field Laboratory, Tallahassee, FL 32310}
\date{\today}

\begin{abstract}
Electron spin resonance (ESR) is usually interpreted as a single-particle phenomenon protected from the effect of many-body correlations. We show that this is not the case in a two-dimensional Fermi liquid (FL) with spin-orbit coupling (SOC). Depending on whether the magnetic field is below or above some critical value, ESR in such a system probes --up to three--collective chiral-spin modes, augmented by the presence of the field, or the Larmor mode, augmented both by SOC and FL renormalizations. We argue that ESR can be used as a probe not only for SOC but also for many-body physics.
\end{abstract}

\maketitle
\emph{Introduction}. Electron Spin Resonance (ESR) spectroscopy is an invaluable tool for studying dynamics of electron spins~\cite{Stormer,Klitzing,ESRcurr}. In a single-particle picture, ESR can be understood either classically,  as resonant absorption of electromagnetic (EM) energy by a classical magnetic moment precessing about the magnetic field, or quantum-mechanically, as absorption of photons with frequency equal to the Zeeman splitting. The absorption rate, $w$, of an incident
electromagnetic (EM) wave (with frequency $\Omega$ and amplitudes of the electric and magnetic fields $\vec{E}^{\text{em}}$ and $\vec{B}^{\text{em}}$) is given by the Kubo formula \cite{Shekhter,CIESR,abs}
\beq\label{eq:1}
w=2\sum_{ij}\left[\sigma'_{ij}(\Omega)E_{i}^{\text{em}}E_{j}^{\text{em}} + \Omega\chi''_{ij}(\Omega)B_{i}^{\text{em}}B_{j}^{\text{em}}\right],
\eeq
where $\sigma_{ij}'(\Omega)$ is the real part of the conductivity and $\chi_{ij}''(\Omega)$ is the imaginary part of the spin susceptibility.

If the static magnetic field ($\vec B$) is in the plane of a two-dimensional electron gas (2DEG) and there is no spin-orbit coupling (SOC), the only resonant feature is due to a pole in the second term of Eq.~(\ref{eq:1}) at the Larmor frequency. This is a conventional (or direct) ESR. However, because the spin susceptibility is proportional to $1/c^{2}$, where $c$ is the speed of light, the direct ESR signal is very weak. SOC of Rashba~\cite{R1,bychkov} and/or Dresselhaus~\cite{Dresselhaus} types changes the situation drastically by producing an effective magnetic field, which acts on the spin of an electron with given momentum $\vec p$ and is proportional to $|\vec p|$. The driving electric field (either from a {\em dc} current or EM wave) gives rise to a flow of electrons with a non-zero drift momentum; hence the electron system as a whole experiences an effective magnetic field due to SOC. The magnitude of bare SOC is strongly enhanced by virtual interband transitions~\cite{Winkler}; as a result, the electric component of an EM field couples to electron spins much stronger than the magnetic one. This is an electric dipole spin resonance (EDSR) \cite{RE,R2,R3,LossEDSR}, which gives rise to a range of spectacular phenomena, e.g., a strong enhancement  of microwave absorption in a geometry when $\vec E^{\text{em}}$ is in the plane of a 2DEG~\cite{EDSR_AlAs} and a shift of the resonance frequency by a {\em dc} current~\cite{CIESR,ESRcurr}.

\begin{figure*}[htp]
$\begin{array}{ccc}
\includegraphics[width=1in]{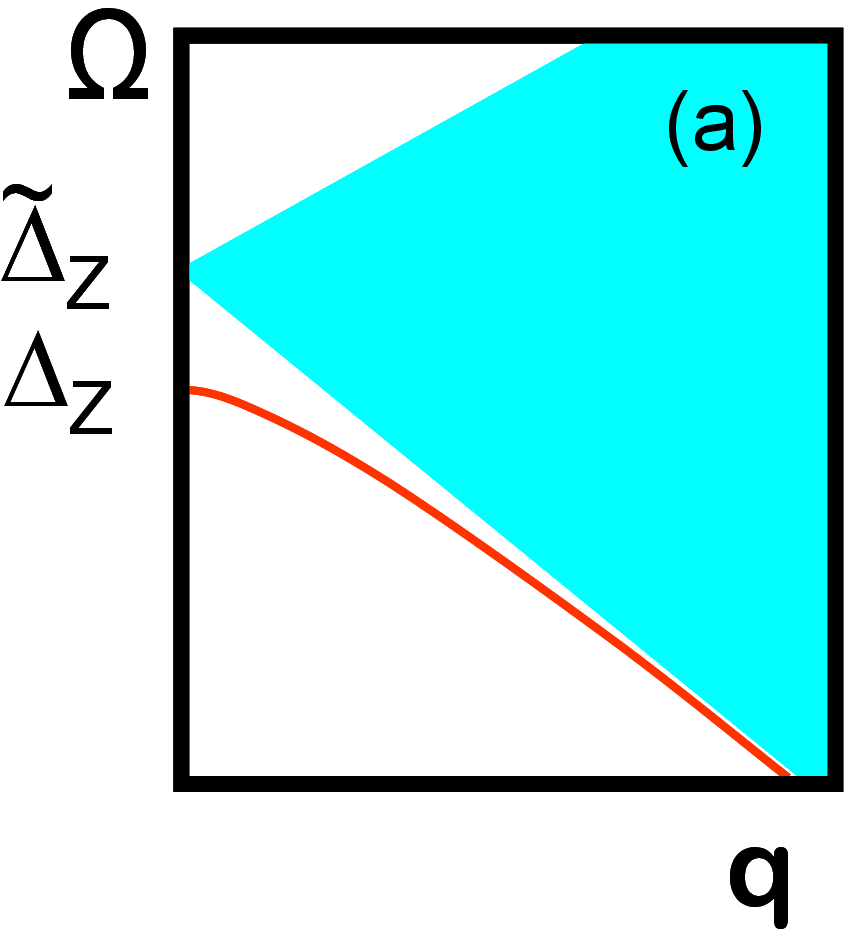}&
\includegraphics[width=2.7in]{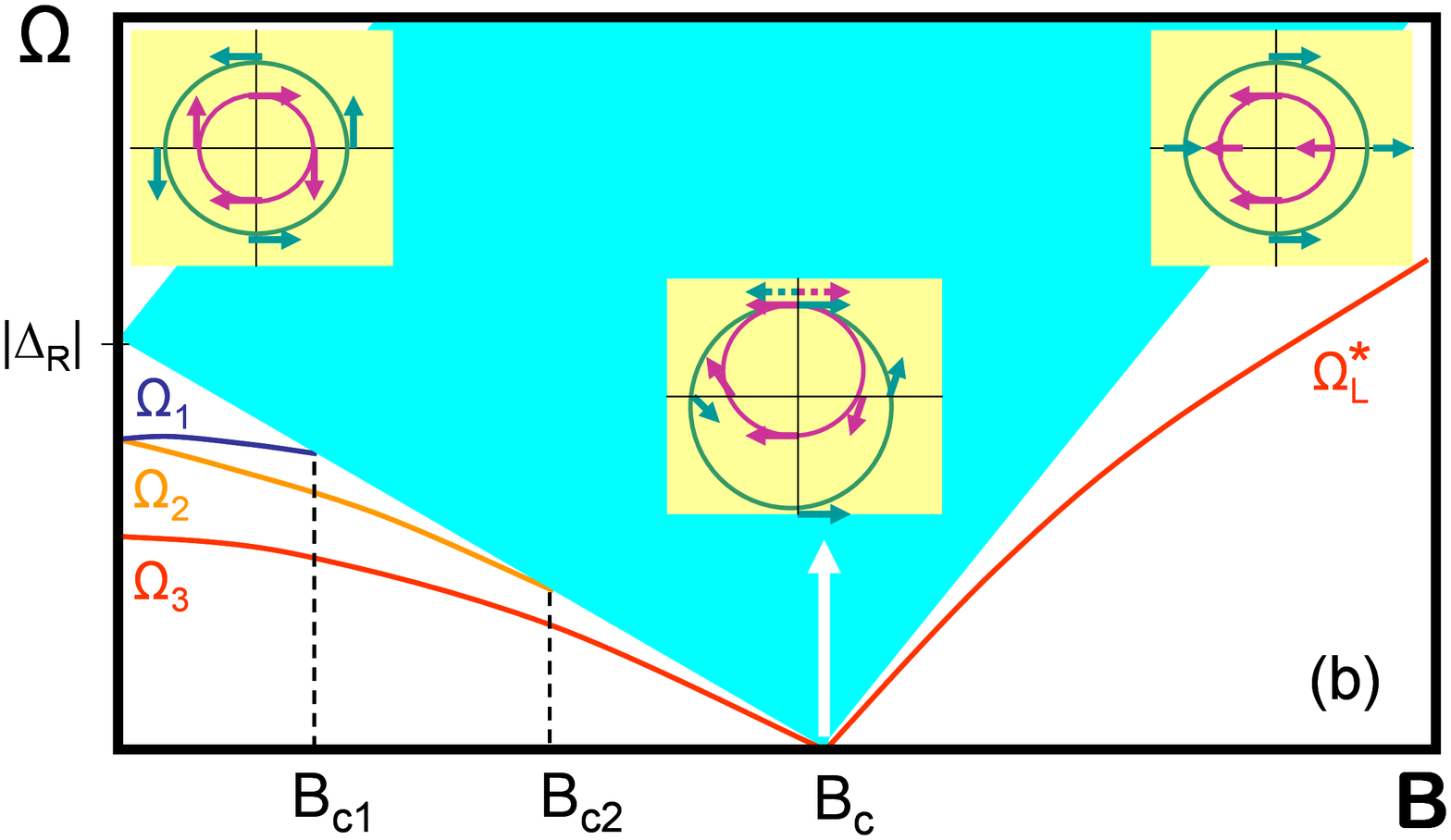}&
\includegraphics[width=3in]{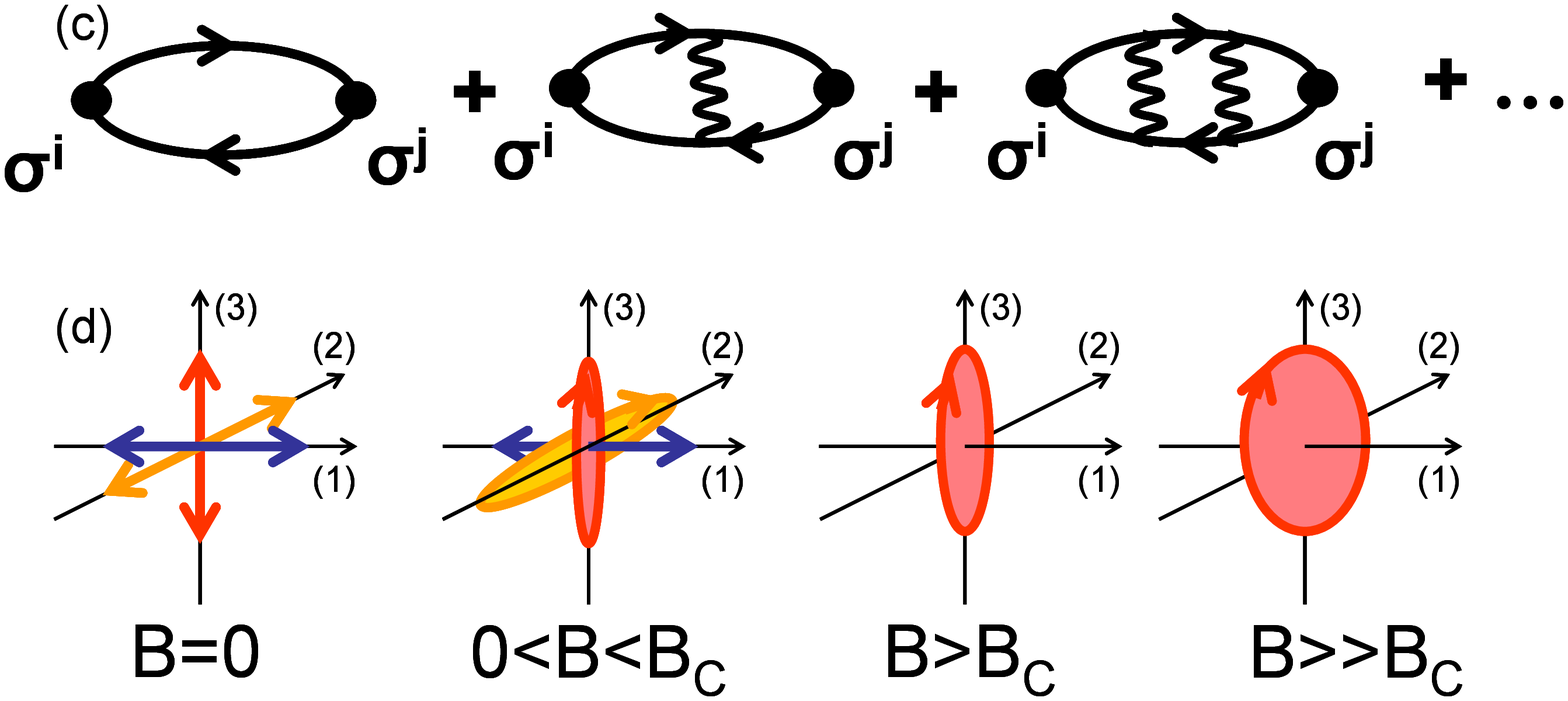}
\end{array}$\caption{\label{fig:schema}[color on-line] (a)
The Silin-Leggett mode (red) and continuum of spin-flip excitations (shaded, blue) for a Fermi liquid in the magnetic field. $\Delta_Z$ and $\bar\Delta_Z$ are the bare and renormalized Zeeman energies, correspondingly. (b) Schematically:  the frequencies of the collective modes and  continuum boundaries as a function of $B$ for a Fermi liquid with Rashba spin-orbit coupling in the magnetic field.  The gap in the continuum closes at the critical field $B_c$, where the spin-split bands become degenerate.  For $B<B_c$, there are three chiral-spin modes, $\Omega_{1\dots 3}$. For $B>B_c$, there is one mode with a renormalized Larmor frequency, $\Omega^*_L$. Insets: spin-split Fermi surfaces.  (c) RPA diagrams for the spin susceptibility. (d) Evolution of polarizations of the collective modes with $B$.}
\end{figure*}

In this Letter, we discuss the effect of the electron-electron interaction on the ESR signal. In the Fermi-liquid (FL) language, ESR in the absence of SOC is an excitation of the Silin-Leggett (spin-flip) collective mode \cite{silin:1958,leggett:1970}, cf. ~Fig.~\ref{fig:schema}a. Although the very existence of this dispersive mode is due to many-body correlations, its end point at $q=0$--the Larmor  frequency--is protected from renormalizations by these correlations and given by the bare Zeeman energy ~\cite{Yafet}. In addition, there is a continuum of spin-flip single-particle excitations (shaded region in Fig.~\ref{fig:schema}a), whose end point corresponds to the renormalized Zeeman energy.  Although the absorption rate should, in principal, contain the contributions  from both the collective mode and continuum, the latter does not contribute to ESR because its spectral  vanishes at $q=0$. These two main features of the ESR signal--no many-body renormalization of the resonance frequency and no contribution from the continuum--are due to conservation of the total spin ($\vec S$) projection onto $\vec B$.

The situation changes drastically in the presence of SOC, which breaks conservation of $\vec S\cdot\vec B$ and thus gives rise to fundamentally new features in the excitation spectrum discussed in this Letter. Depending on whether the ratio of the Zeeman energy to spin-orbit splitting is larger than, comparable with, or smaller than unity, one can define the regimes of ``high'', ``moderate", and ``weak'' magnetic fields. We show that the ESR frequency in the high-field regime is affected both by SOC and many-body correlations and scales non-linearly with $B$ (see Fig.~\ref{fig:schema}b). The deviation from linearity can be used to extract the amplitudes of both SOC and electron-electron correlations. In addition to the resonance peak, the ESR signal now also shows a broad feature due the continuum of spin-flip excitations. In the presence of SOC, the resonance itself is entirely a many-body effect; in the absence of interactions, the signal comes entirely from the continuum~\cite{Starykh}. The conventional ESR regime
is reached in the limit of $B\to\infty$. As the field gets weaker, the ESR frequency scales down and finally vanishes at a critical field, $B_c$, where the spin-split energy levels become degenerate (see insets in Fig.\ref{fig:schema}b) and the gap in the continuum closes. The region around $B_c$ defines the moderate-field range.
For $B<B_c$, the resonance appears again and two more modes split off the continuum as the field passes through the critical values, $B_{c2}$ and $B_{c1}$. At $B\to 0$, the three modes evolve into chiral-spin resonances--collective oscillations of magnetization in the absence of the magnetic field \cite{Shekhter,Ali1,SM1}. In the most general case of both Rashba and Dresselhaus SOC present, all the three chiral-spin modes are ESR-active.

In the prior literature, the discussion of the effect of SOC on ESR was limited to two aspects: D'yakonov-Perel' damping \cite{DP} of the signal and coupling of electron spins to the electric field via the EDSR mechanism. We show in this Letter that the effect of SOC is much richer than the two aspects mentioned above. To the best of our knowledge, all the experiments thus far have been performed in the high-field limit, where the effect of SOC is quantitative rather than qualitative. We propose to study ESR in moderate and weak field regimes, where the SOC-induced changes are qualitative.

\emph{Model and Formalism}. We study a two-dimensional (2D) electron system with both Rashba and Dresselhaus types of SOC (RSOC and DSOC, correspondingly) and in the presence of an in-plane magnetic field. We adopt the form of Dresselhaus SOC appropriate for a GaAs [001] quantum well and choose the $x_1$ and $x_2$ axes along
the $[1\bar 10]$ and $[110]$ directions, correspondingly. The single-particle part of the Hamiltonian then reads \cite{Ganichev}
\bea\label{eq:ham}
\hat{\mathcal{H}}_0&=&\frac{\bk^2}{2m}\hat\sigma_0 + \alpha\left(\hat\sigma_1k_2-\hat\sigma_2k_1\right)\nonumber\\
&&+\beta\left(\hat\sigma_1
k_2+\hat\sigma_2k_1\right) - \frac{g\mu_B}{2}\hat{\sigma}_1{B},
\eea
where $m$ is the band mass, $\mu_B$ is the Bohr magneton, $\hat\sigma_{1,2,3}$ are the Pauli matrices, $\hat\sigma_0$ is the $2\times2$ identity matrix, and $\alpha$ ($\beta$) is the Rashba (Dresselhaus) parameter. For simplicity, we chose the magnetic field to be along one of the two high-symmetry directions.  This restriction will be relaxed later on.

The many-body part of the Hamiltonian, $\hat{\mathcal H}_{\text{int}}$, depends only on the electron positions, $\hat{\vec x}$. Consequently,  $[\hat{\mathcal H}_{\text{int}},\hat{\vec x}] =0$ and the velocity operator $\hat {\vec v}=i[\hat{\mathcal H}_0+\hat{\mathcal H}_{\text{int}},\hat{\vec x}]=i[\hat{\mathcal H}_0,\hat{\vec x}]$ retains its free-electron form:
\beq
\hat {\vec v}=\left(\frac{k_1}{m}\hat\sigma_0-(\alpha-\beta)\sigma_2
,\frac{k_2}{m}\hat\sigma_0+
(\alpha+\beta)\sigma_1
\right).
\eeq
The gradient terms in $ \hat {\vec v}$ give rise to the Drude part of the conductivity, while the spin-dependent terms give rise to its $B$-dependent part, $\sigma_B$, which determines the EDSR signal. In the Voigt geometry ($\vec E^{\text{em}}||\vec B\perp\vec B^{\text{em}}$), the first (EDSR) term in the absorption rate [Eq.~(\ref{eq:1})]  contains the component $(\sigma'_B)_{11}$, which is related to the spin susceptibility via
\beq (\sigma'_B)_{11}=\frac{e^2}{(g\mu_B)^2\Omega}(\alpha-\beta)^2\chi_{22}'',
\label{eq:dm1}
\eeq
while the second (ESR) term contains $\chi''_{22/33}$ for $\vec B^{\text{em}}||\hat x_{2/3}$. Equation (\ref{eq:dm1}) also holds in the presence of the electron-electron interaction. The ratio of the ESR to EDSR terms  is controlled by a small parameter $(\lambdabar_C/\lambdabar_F)^2\sim 10^{-9}-10^{-8}$, where $\lambdabar_F$ is the Fermi wavelength and $\lambdabar_C=\hbar/mc$ is the Compton length~\cite{Shekhter}. Therefore, $w$ is determined by $(\sigma'_B)_{11}$ to very high accuracy.

We assume that both the spin-orbit splitting and Zeeman energy are much smaller than the Fermi energy. In this case, the corresponding terms in the Hamiltonian can be treated as perturbations to the conventional, SU(2)-invariant FL,  and complications encountered in generalizing the FL theory for arbitrarily large spin-dependent terms~\cite{Meyerovich,Ali2} do not arise. The ESR signal is completely characterized by the spin susceptibility. At $q=0$, the spin and charge sectors of the theory decouple because of charge conservation~\cite{SM1},
and $\chi_{ij}(\Omega)$ can be found within the usual random-phase approximation (RPA) shown in Fig.~\ref{fig:schema}c, where the Green's functions include the $B$-dependent shifts of the chemical potential
\bea\label{eq:chi}
\chi_{ij}(\Omega_m)&=&-\frac{(g\mu_B)^2}{4}\Pi^0_{ij'}(\Omega_m)\left[1+\frac U2\hat\Pi^0
(\Omega_m)\right]^{-1}_{j'j},
\eea
where $\Pi^0_{ij}(\Omega_m)=\int_K\text{Tr}\left[\hat{\sigma}_i\hat{G}_K\hat{\sigma}_j\hat{G}_{K+Q}\right]
$
with
$i,j\in\{1,2,3\}$, $Q=(i\Omega_m,\vec{0})$; $K=(i\omega_m,\vec{k})$, and $\int_K\equiv T\sum_{\omega_m}\int\frac{d^2k}{(2\pi)^2}$.
Furthermore, $\hat{G}^{-1}_K=(i\omega+\mu)\hat{\sigma_0} -
\hat{\mathcal{H}}'_0 $, where $
\hat{\mathcal{H}}'_0$ differs from $\hat{\mathcal{H}}_0$ in that the Zeeman energy is replaced by its renormalized value : $g\mu_BB\rightarrow g\mu_BB/(1-u)$,
where $u\equiv \nu_{2D}U$ and $\nu_{2D}=m/2\pi$ is the density of states in 2D \cite{supp_FL2}.
For weak SOC, i.e., for $\alpha$, $\beta\ll v_F$ with $v_F$ being the Fermi velocity in the absence of SOC,
the system is characterized by four energy scales:
\beq\label{eq:scales}
\Delta_R\equiv 2\alpha k_F;~\Delta_D\equiv 2\beta k_F;~\Delta_Z\equiv g\mu_BB;~
\tilde{\Delta}_Z=\frac{\Delta_Z}{1-u},
\eeq
where $k_F=mv_F$. We choose the Zeeman energies to be positive, while the signs of $\Delta_R$ and $
\Delta_D$ are arbitrary.

\emph{ESR without SOC}. We start by revisiting the well-known case of a FL without SOC in the magnetic field ($\alpha=\beta=0$ in Eq. \ref{eq:ham}). In this case,$\Pi_{1j}^0(\Omega_m)=0$ ($j\in\{1,2,3\}$) because the projection of spin on the direction of $\vec B$ is conserved. For the rest of the components we obtain, upon analytic continuation ($i\Omega_m\to \Omega+i0^+$):
$\Pi^0_{22}(\Omega)=\Pi^0_{33}(\Omega)=2\nu_{2D}\tilde{\Delta}_Z^2/(\Omega^2-\tilde{\Delta}^2_Z)$ and
$\Pi^0_{23}(\Omega)=-\Pi^0_{32}(\Omega)=-2i\nu_{2D}\Omega\tilde{\Delta}_Z/(\Omega^2-\tilde{\Delta}^2_Z)$. The collective mode corresponds to a pole of Eq.~(\ref{eq:chi}), when det$\left[1+\frac U2\Pi^0_{ij}(\Omega)\right]=0$ or $1+\frac U2\Pi^0_{22} = \pm \frac U2 i\Pi^0_{23}$. The only solution of this equation outside the spin-flip continuum is the Larmor frequency: $\Omega_L=\tilde{\Delta}_Z(1-u)=\Delta_Z$. On the other hand, $\chi_{ij}''(\Omega)$ vanishes at the continuum ($\Omega=\tilde\Delta_Z$), and thus the continuum does not contribute to ESR.

\begin{figure*}[htp]
\includegraphics[width=17cm]{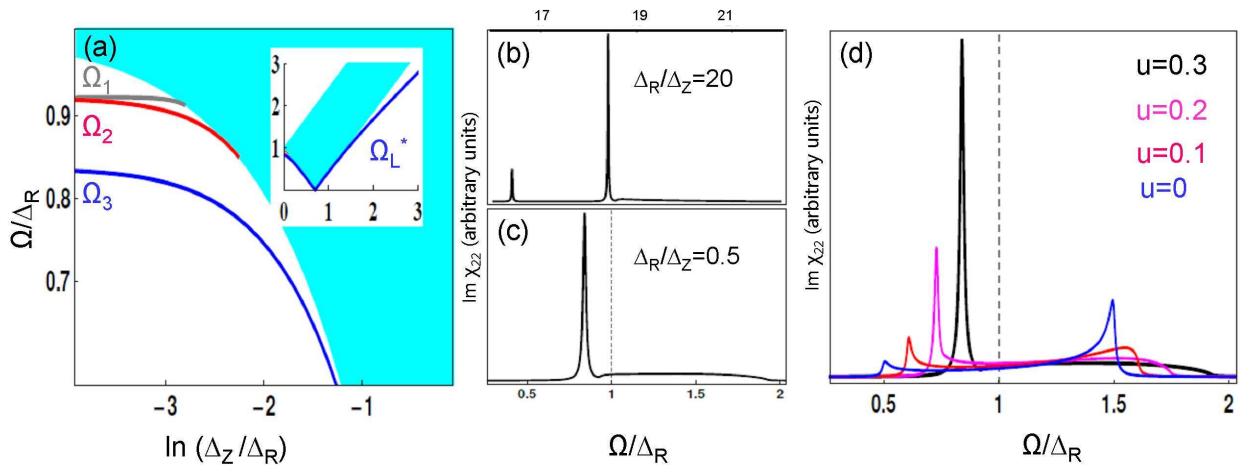}
\caption{[color on-line] (a) Chiral-spin modes as a function of the Zeeman energy, $\Delta_Z$, in units of the Rashba spin splitting, $\Delta_R$ (on a semi-logarithmic scale). Inset: renormalized Larmor mode ($\Omega^*_L$) at higher fields. (b) Imaginary part of the susceptibility in the weak field limit ($\Delta_R/\Delta_Z=20$).
(c) Same as in the high-field limit. The dashed line marks the bare Larmor frequency ($\Omega_L$). The continuum is seen as a broad hump to the right of the resonance. In panels (a)-(c), the dimensionless interaction is $u=0.3$. (d) Evolution of the ESR signal with $u$.  Damping of $\Gamma=0.01\Delta_R$ was added to the Green's functions to mimic the effect of disorder in all plots.
\label{fig:3}}
\end{figure*}

\emph{ESR with RSOC}.
This case is realized by setting $\beta=0$ in Eq.~(\ref{eq:ham}). After including the self-energy correction to the Zeeman term~\cite{supp_FL2,comment_alpha}, the dispersions of the spin-split bands become
$\e^{\pm}_{\bk}=k^2/2m \pm\frac12\sqrt{(2\alpha k)^2 + (\tilde{\Delta}_Z)^2 - 2\tilde{\Delta}_Z(2\alpha k)\sin\theta_k}$, where $\theta_k$ is the angle between $\bk$ and the $x_1-$axis. Although the spin projection onto $\vec B$ is not conserved anymore, some off-diagonal components of $\Pi_{ij}^0$ still vanish. Indeed, since $\vec{B}^{\text{em}}\times\vec B=0$ for $\vec{B}^{\text{em}}\parallel \vec B\parallel \hat x_1$, the only two pseudo-vectors in the system are $\vec{B}^{\text{em}}$ and $\vec B$ themselves. The magnetization induced by $\vec{B}^{\text{em}}$ is  also a pseudo-vector and thus can only be parallel to $\vec B$, which implies that $\Pi^0_{1j}=0$ for $j=2,3$. The non-zero components of  $\hat\Pi^0$ are given by~\cite{supplement_3}:
\bea
\Pi^0_{11}(\Omega)&=&-2\nu_{2D}\frac{W^2(1-f)}{4\tilde{\Delta}_Z^2},\label{eq:piRB}\\
\Pi^0_{22}(\Omega)&=&-2\nu_{2D}\left[\frac{\tilde{\Delta}_Z^2}{fW^2}+\left(1-\frac1f\right)\left(1-\frac{W^2}{4\tilde{\Delta}_Z^2}\right)\right],\nonumber\\
\Pi^0_{33}(\Omega)&=&-2\nu_{2D}\left[1+\frac{\Omega^2}{fW^2}\right],\nonumber\\
\Pi^0_{23}(\Omega)&=&2\nu_{2D}\frac{i\Omega}{\tilde{\Delta}_Z}\left[\frac12\left(1-\frac1f\right)+ \frac{\tilde{\Delta}_Z^2}{fW^2}\right]=-\Pi^0_{32}(\Omega),\nn
\eea
where $f\equiv\sqrt{1-4\Delta_R^2\tilde{\Delta}_Z^2/W^4}$ and $W^2\equiv  \Delta_R^2 + \tilde{\Delta}_Z^2-\Omega^2-i0^+\text{sgn}\Omega$. The formulas above reduce to the known limits ~\cite{SM1} when $\Delta_R\rightarrow 0$ and $\Delta_Z\rightarrow0$, respectively.

The band energies vary around the Fermi surface, reaching the maximum and minimum values of $\left|\tilde\Delta_Z\pm\left|\Delta_R\right|\right|$, correspondingly. As a result, the continuum of spin-flip excitations occupies a finite interval of frequencies $\left|\tilde{\Delta}_Z-|\Delta_R|\right|<\Omega<\tilde{\Delta}_Z+|\Delta_R|$, where all $\Pi^0$'s in Eq.~(\ref{eq:piRB}) have non-zero imaginary parts. This is in contrast to the case of $\alpha=0$, where the continuum has zero spectral weight (see Fig. \ref{fig:schema}a). The gap in the continuum closes at a special field, $B_c$, such that $\tilde{\Delta}_Z(B_c)=|\Delta_R|$ and the spin-split bands become degenerate (Fig. \ref{fig:schema}b).

The collective modes correspond to the poles of Eq.~(\ref{eq:chi}) outside the continuum. The eigenmode equation splits into two:
\begin{subequations}
\bea
1+\frac{U}{2}\Pi^0_{11}(\Omega)&=&0,\label{eq:2a}\\
\left[1+\frac{U}{2}\Pi^0_{22}(\Omega)\right]\left[1+\frac{U}{2}\Pi^0_{33}(\Omega)\right]&=&-\frac{U^2}{4}\left[\Pi^0_{23}(\Omega)\right]^2.\nn\\
\label{eq:2b}
\eea
\end{subequations}
For $B>B_c$, Eq.~(\ref{eq:2a}) has no solutions while Eq.~(\ref{eq:2b}) has a unique solution~\cite{supplement_4}, which is the Larmor frequency, $\Omega_L^*$, renormalized both by SOC and electron-electron interaction (cf. inset in Fig.~\ref{fig:3}a). At the highest fields ($\tilde{\Delta}_Z\gg|\Delta_R|/u$),
\beq\label{eq:larmor1}
\Omega^*_L\approx\Delta_Z\left[1-\frac{(2-3u)(1-u)}{4u}\left(\frac{\Delta_R}{\Delta_Z}\right)^2\right].
\eeq
When $B$ is just slightly above $B_c$, i.e., $\tilde{\Delta}_Z
\approx|\Delta_R|$ but still $\tilde\Delta_Z>|\Delta_R|$, we get
\beq\label{eq:larmor2}
\Omega^*_L\approx\left(
\tilde\Delta_Z-|\Delta_R|
\right)\left[1-\frac{u^2(1-\frac{3u}{4})^2}{2(1-\frac{u}{2})^2(1-u)^2}\frac{(\tilde\Delta_Z-|\Delta_R|)^2}{\tilde\Delta^2_Z}\right].
\eeq
In the limit of $u\ll1$, we have an additional regime defined by $\Delta_R\ll\tilde{\Delta}_Z\ll\Delta_R/u$, where
\beq
\Omega^*_L\approx|\tilde\Delta_Z|\left[1-\frac{u^2\tilde\Delta_Z}{2|\Delta_R|}
\right].
\eeq

For $B<B_c$, Eq.~(\ref{eq:2a}) has one solution, $\Omega=\Omega_1$, which corresponds to oscillations of the $x_1$ component of the magnetization $\vec M$, while Eq.~(\ref{eq:2b}) has two solutions, $\Omega=\Omega_2$ and $\Omega=\Omega_3$, which correspond to coupled oscillations of the components $M_2$ and $M_3$. The $\Omega_1$ and $\Omega_2$ modes run into the continuum at fields $B_{c1}$ and $B_{c2}$, correspondingly (cf.~Fig.~\ref{fig:schema}b). The three modes are plotted in Fig.~\ref{fig:3}a for a range of fields below $B_c$. As the field is lowered further, these three solutions evolve into the chiral-spin resonances~\cite{Shekhter,Ali1}. At $B=0$, $\Pi_{23}^0$ in Eq.~(\ref{eq:2b}) vanishes by time-reversal symmetry, while $\Pi^{0}_{11}$ and $\Pi_{22}^0$ become equal by the $C_{\infty v}$ symmetry. In this limit, $\Omega_1=\Omega_2=|\Delta_R|\sqrt{1-u/2}$ and $\Omega_3=|\Delta_R|\sqrt{1-u}$~\cite{SM1}.

In the absence of  DSOC, absorption is determined entirely by $\chi_{22}''$ [cf.~Eq.~(\ref{eq:dm1})]. Since the $\Omega_1$ mode is decoupled from the $\Omega_2$ and $\Omega_3$ modes, it is ESR-silent. The magnetic field couples the $\Omega_2$ and $\Omega_3$ modes, both of which show up in ESR. For $B>B_c$, there is only one ESR-active mode, whereas for $B<B_c$ there can be one or two active modes, depending of whether $B$ is smaller or larger than $B_{c2}$. In addition to a sharp peak(s), there is also a broad feature corresponding to absorption by the continuum of spin-flip excitations.

Figure \ref{fig:3}d depicts the evolution of the ESR signal with increasing $u$. In the presence of SOC, a sharp mode occurs only due to many-body interaction, as it pushes the mode away from the continuum. This is in contrast to the case without SOC, where the mode exists even without interaction. Both the peak and broad hump due to the continuum--have been observed in Ref.~\cite{CdMnTe}, although the detailed shape of the hump is yet be explained.

As the magnetic field increases from zero to values exceeding $B_c$, polarization of the collective modes changes qualitatively (cf.~Fig.~\ref{fig:schema}d).  At $B=0$,
the susceptibility is diagonal, which means that the different components of the magnetization oscillate independently and are thus linearly polarized.  For $0<B<B_c$, the $M_1$ component is still linearly polarized, while coupled oscillations of the $M_2$ and $M_3$ components can be decomposed into two elliptically polarized modes. For $B>B_c$, there is only one elliptically polarized mode which evolves into a circularly polarized Larmor mode for $B\gg B_c$.

{\it ESR with both RSOC and DSOC}.
Adding DSOC to RSCO lowers the symmetry from $C_{\infty v}$ to  $C_{2v}$.
As a result, the doubly degenerate chiral-spin resonance splits into two already at $B=0$. Other than that, DSOC does not change the situation qualitatively, as long as $\vec B$ is along the high-symmetry axis [as in Eq.~(\ref{eq:ham})] :  one of the three modes is still ESR-silent, so the signal consists of up to two lines. If $\vec B$ is along a generic in-plane direction, all modes become ESR-active, and the signal consists of up to three lines \cite{supplement_3}.

The drastic modification of the ESR signal by SOC should be seen in most conventional quantum wells \cite{Sup}, given that the microwave frequency and the magnetic field are properly tuned. Recent advances in microwave technology \cite{ESRrange} have greatly broadened the range of frequency tuning. The sensitivity of the Larmor mode to many-body interactions in the presence of SOC suggests that one has to account for many-body effects when  extracting the $g$-factors and SOC parameters from the precession measurements \cite{VS1,VS2,VS4}. One of the open questions is whether the chiral-spin modes can be detected through electrically detected ESR \cite{RB2,RB1,EdESR,EdESR2}.


We would like to thank C. R. Bowers, I. Paul, F. Perez, E. I. Rashba, and C. A. Ullrich for useful discussions. SM acknowledges the Dirac Fellowship award from the NHMFL, which is supported by the NSF via Cooperative agreement No. DMR-1157490, the State of Florida, and the U.S. DoE. DLM acknowledges support from the NSF via grant NSF DMR-1308972 and Stanislaw Ulam Scholarship at the CNLS, LANL.

\newpage
\begin{widetext}
\appendix
\section{Single-particle Hamiltonian: eigenstates and self-energy correction}
The Hamiltonian in Eq.~(2) of the main text (MT) can be written as
\bea\label{eq:app1}
\hat{\mathcal{H}}_0 &=& \frac{\vec{k}^2}{2m}\hat\sigma_0 + \lambda_{k,\theta} k \left(\sin\phi_k \hat{\sigma}_1 - \cos\phi_k \hat{\sigma}_2\right),
\eea
where the parameters $\lambda$ and $\phi_k$ are defined by the following relations
\bea
2\lambda_{k,\theta} k &=& \sqrt{(2\alpha k)^2 + (2\beta k)^2 - 8\alpha \beta k^2 \cos2\theta_k + (g\mu B)^2 - 4(g\mu B)(\alpha  + \beta )k\sin\theta_k},\nonumber\\
\sin\phi_k&=&\frac{\alpha+\beta}{\lambda}\sin\theta_k-\frac{g\mu B}{2\lambda k},\nonumber\\
\cos\phi_k&=&\frac{\alpha-\beta}{\lambda}\cos\theta_k.
\eea
The eigenvalues and eigenvectors are given by
\bea\label{eq:app2}
&&\e^{\pm}=\frac{k^2}{2m}\pm \lambda k,\\
&&v^{\pm}=\frac{1}{\sqrt{2}}\left(
\begin{array}{c}
1\\
\mp
ie^{i\phi_k}
\end{array}
\right).
\eea

To account for renomalization of the Zeeman energy entering the Green's function and for possible renormalizations of the spin-orbit parameters, one needs to the find the momentum- and frequency-independent part of the self-energy, $\hat\Sigma$. For an $s-$wave interaction ($U=\text{const}$), $\hat\Sigma$ can be found in the self-consistent Born approximation as (in the notation of the MT):
\bea\label{eq:app3}
&&\hat{\Sigma
}=-U\int_K \hat{G}_{K},~~~~~\hat{G}_{K}=\left(\left[\hat{G}^0_{K}\right]^{-1}-\hat{\Sigma} \right)^{-1},
\eea
where $[\hat{G}^0_K]^{-1}=(i\omega_m+\mu)\hat{\sigma}_0-\hat{\mathcal{H}}_{0}$. By construction, $\hat\Sigma$ does not depend on $K$
and thus can be written as
\beq
\hat{\Sigma}= \sum_{\ i=1\dots3}a_{i}\hat{\sigma}_{i},
\eeq
where the coefficients $a_{i}$ are to be determined. Note that we dropped the coefficient $a_0$ as it would only result in a shift of the chemical potential. Solving the algebraic matrix equation, we get $a_{1}=\frac{u}{1-u}\frac{g\mu B}{2}$ (where $u\equiv \frac{mU}{2\pi}$), and $a_{2}=a_{3}=0$. This amounts to changing $g\mu B\rightarrow \frac{g\mu B}{1-u}$ or $\Delta_Z\rightarrow \tilde\Delta_Z$ as in the MT. Since $a_{2}=a_{3}=0$, the spin-orbit parameters are not renormalized. This is a special feature of the $s$-wave interaction approximation.

The Green's function (with the self-energy correction) is then explicitly written as:
\beq\label{eq:app4}
\hat{G}_{K}=\sum_{r\pm}g_{K}^{r}\hat\Omega_r,~~~~\hat\Omega_r=\frac12\left[\hat\sigma_0+ r(\hat\sigma_1\sin\phi_k - \hat\sigma_2\cos\phi_k)\right],
\eeq
where $g_{K}^{r} =1/(i\omega_m -\tilde\e^r_{\bk})$ and $\tilde\e^r_{\bk}$ is the electron dispersion which contains the renormalized Zeeman energy: $\Delta_Z\rightarrow \tilde\Delta_Z$.

\section{Collective modes within the random-phase approximation}

In this section, we provide some details of analysis of the collective modes within the random-phase approximation (RPA).

\subsection{General case: collective modes in the presence of the magnetic field, and both Rashba and Dresselhaus spin-orbit couplings}
Within the RPA framework, one needs to find all components of the spin-charge polarization tensor $\Pi^0_{ij}$. This is a challenging task in the most general case, when the magnetic field and both Rashba and Dresselhaus types of spin-orbit coupling (RSCO and DSOC, correspondingly) are present.  However, in the limit of weak both magnetic field and SOC, i.e. when the Zeeman energy and spin-orbit splitting of the energy bands are small compared to the Fermi energy, one can confine the momentum integration to the vicinity of the Fermi surface.
In this section, we choose the magnetic field to be along an arbitrary in-plane direction. Consequently, the Zeeman term in Eq.~(2) of MT is replaced by $-(g\mu_B/2)\left(\hat\sigma_1B_1+\hat\sigma_2B_2\right)$. The corresponding changes in the eigenvalues and eigenvectors can readily be traced down; we will refrain from giving explicit forms here.  Replacing the integration over the radial component of the momentum by that over the electron dispersion, we arrive at two types of integrals [as in the MT, $Q$ stands for the $2+1$ bosonic momentum with zero spatial part: $Q=(i\Omega_m,0)$] :\bea\label{eq:type1}
\frac12\int d\e (g^+_Kg^-_{K+Q}+g^-_Kg^+_{K+Q}) &=& \frac12\int d\e\left\{\frac{n_F(\e^+)-n_F(\e^-)}{i\Omega_m + \e^+-\e^-} ~+ ~(+\rightarrow -)\right\},\nonumber\\
&=&-\frac{\Lambda_\theta^2}{\Omega_m^2+\Lambda_\theta^2}
\eea
and
\bea\label{eq:type1}
\frac12\int d\e (g^+_Kg^-_{K+Q}-g^-_Kg^+_{K+Q})&=&\frac{i\Omega_m\Lambda_\theta}{\Omega_m^2+\Lambda_\theta^2},
\eea
where
$\Lambda_\theta\equiv 2\lambda_{k=k_F,\theta}k_F$ is the SOC splitting at point $\theta$ on the Fermi surface, $\lambda_{k,\theta}$
is obtained from Eq.~(\ref{eq:app1}) by adding the second component of the magnetic field (which results in $\cos\phi_k\rightarrow (\alpha-\beta)\cos\theta_k/\lambda+g\mu B/(2\lambda k)$), and $k_F$ is the Fermi  momentum in the absence of both the magnetic field and SOC. Using these relations, we get
\bea\label{eq:rdb}
\Pi^0_{11}(\Omega_m)&=&-2\nu_{2D}\int\frac{d\theta}{2\pi}\frac{\Lambda_\theta^2}{\Omega_m^2+\Lambda_\theta^2}\cos^2\phi_k,\nonumber\\
\Pi^0_{12}(\Omega_m)&=&-\nu_{2D}\int\frac{d\theta}{2\pi}\frac{\Lambda_\theta^2}{\Omega_m^2+\Lambda_\theta^2}\sin2\phi_k,\nonumber\\
\Pi^0_{21}(\Omega_m)&=&\Pi^0_{12}(\Omega_m),\nonumber\\
\Pi^0_{13}(\Omega_m)&=&2\nu_{2D}\int\frac{d\theta}{2\pi}\frac{\Omega_m\Lambda_\theta}{\Omega_m^2+\Lambda_\theta^2}\cos\phi_k,\nonumber\\
\Pi^0_{31}(\Omega_m)&=&-\Pi^0_{13}(\Omega_m),\nonumber\\
\Pi^0_{22}(\Omega_m)&=&-2\nu_{2D}\int\frac{d\theta}{2\pi}\frac{\Lambda_\theta^2}{\Omega_m^2+\Lambda_\theta^2}\sin^2\phi_k,\nonumber\\
\Pi^0_{23}(\Omega_m)&=&2\nu_{2D}\int\frac{d\theta}{2\pi}\frac{\Omega_m\Lambda_\theta}{\Omega_m^2+\Lambda_\theta^2}\sin\phi_k,\nonumber\\
\Pi^0_{32}(\Omega_m)&=&-\Pi^0_{23}(\Omega),\nonumber\\
\Pi^0_{33}(\Omega_m)&=&-2\nu_{2D}\int\frac{d\theta}{2\pi}\frac{\Lambda_\theta^2}{\Omega_m^2+\Lambda_\theta^2}.
\eea
For simplicity, we consider the magnetic field to be at $45\deg$ to the $x_1$ axis, i.e., $B_1=B_2\equiv B$. Solutions of the eigenmode equation det($1+(U/2)\hat{\Pi}^0$)$=0$ are shown in the left panel of Fig.~\ref{fig:RDB}. In general, there are no qualitative differences compared to the case of only RSOC and the magnetic field, considered in MT: for $B<B_c$ there are two or three modes depending on the ratio $\alpha/\beta$, whereas for $B>B_c$ there is only one mode.
For $\alpha/\beta=0.25$, as chosen in the left panel of Fig.~\ref{fig:RDB}, there are only two modes.
\begin{figure}[htp]
$\begin{array}{cc}
\includegraphics[width=2.95in]{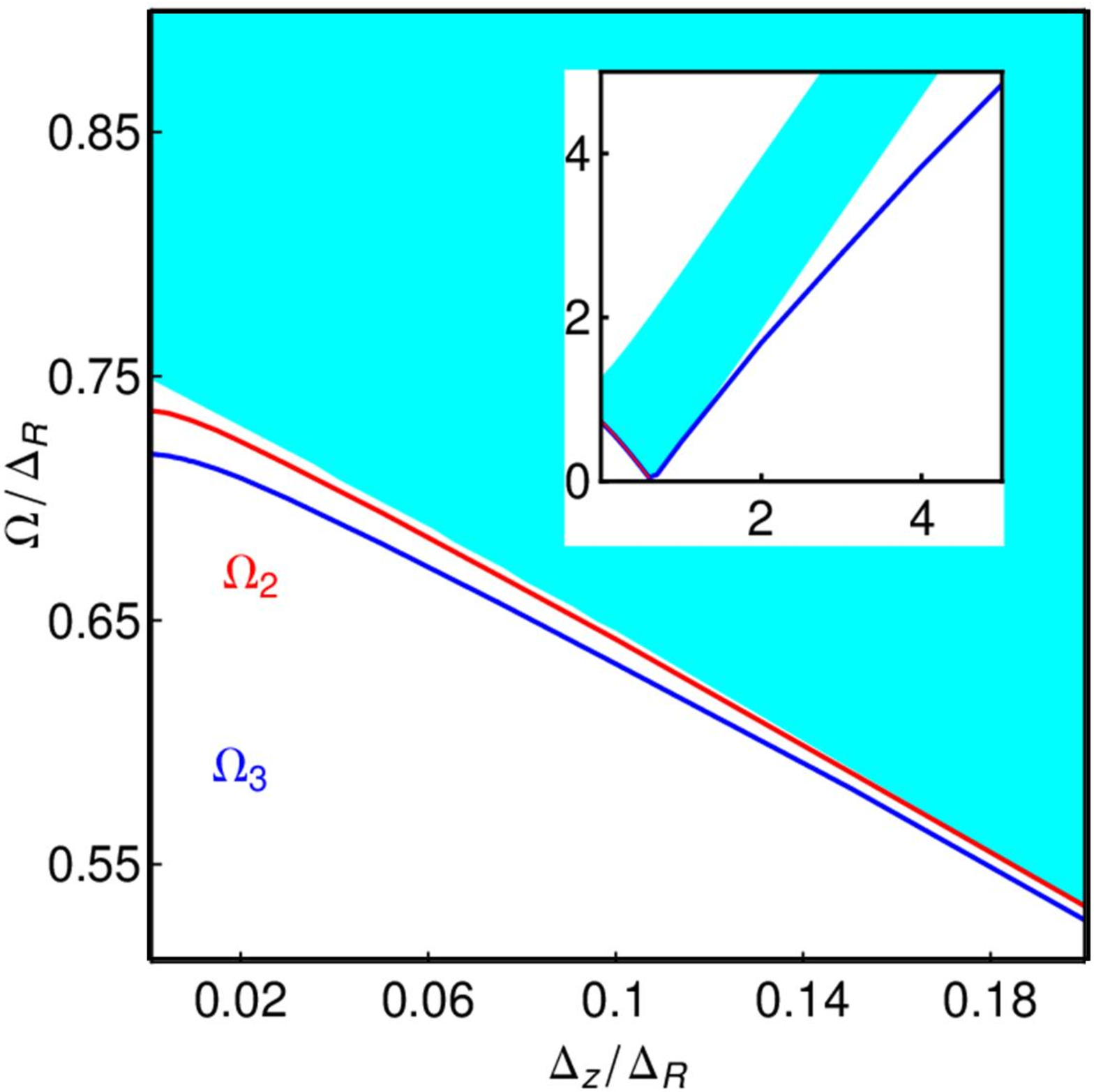}&
\vspace{-0.5cm}\includegraphics[width=2.9in]{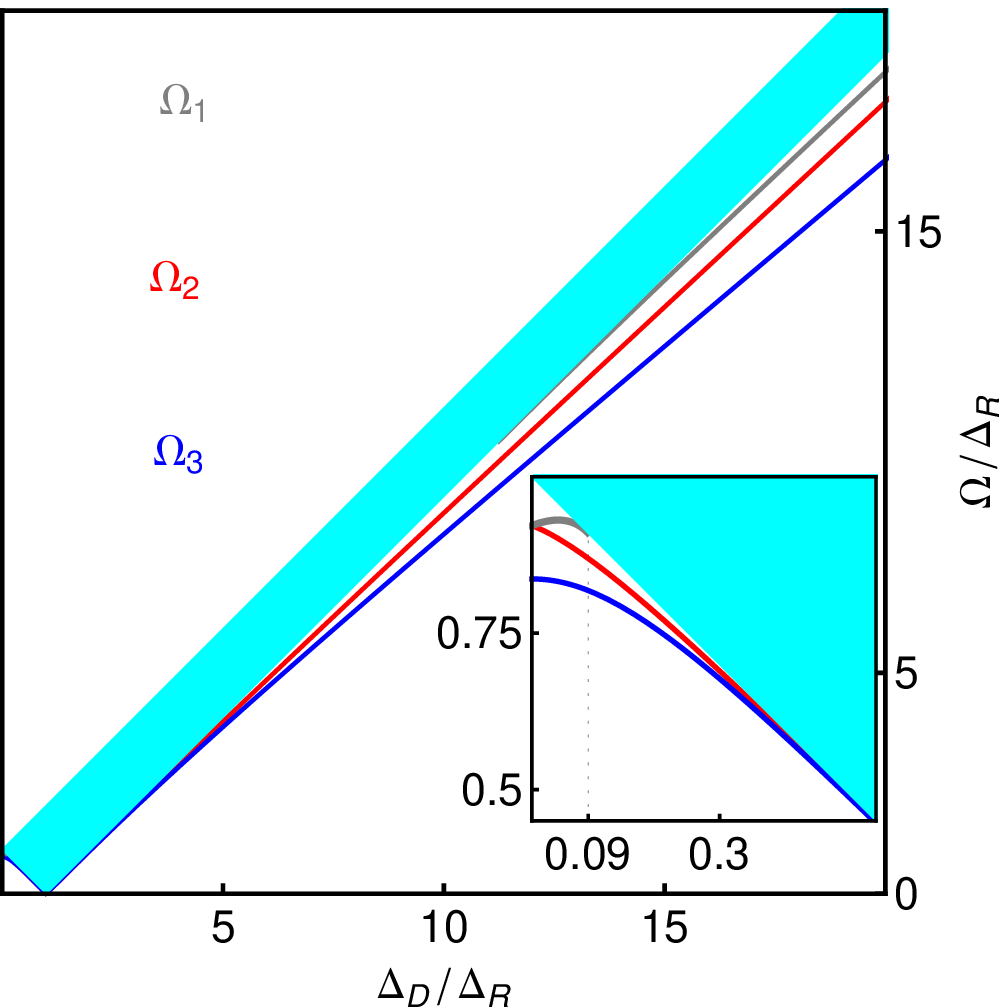}
\end{array}$
\caption{Left: Collective modes in the presence of the magnetic field, and both Rashba and Dresselhaus spin-orbit coupling. $\beta/\alpha=-0.25$. Below the field
at which the gap in the continuum closes, there are two chiral-spin modes; above this field there is only one precessing mode. All modes in this case are elliptically polarized. Right: Collective modes in the presence of Rashba and  and Dresselhaus spin-orbit coupling but in the absence of the magnetic field. There are three modes on either side of the gap closing point. The entire structure of the collective mode is symmetric under $\alpha\rightarrow\beta$. All the three modes are linearly polarized. $u=0.3$ in both plots.
\label{fig:RDB}}
\end{figure}
\subsection{Collective modes in zero field with Rashba and Dresselhaus spin-orbit coupling}
This scenario provides the reference point ($B=0$) to study the magnetic field dependence of the chiral-spin modes. We will show here that in the absence of $B$, there are only linearly polarized oscillations of magnetization. This scenario also highlights the differences and similarities between adding DSOC or the magnetic field to RSOC. In the former case, the crystalline lattice reduces the symmetry from $C_{\infty v}$ to $C_{2v}$, while in the latter case rotational symmetry is completely lost. This actually has important consequences, as shall become apparent. In the case of $\alpha\neq 0$, $\beta\neq 0$, $B=0$, the spin-flip continuum occupies  the energy interval $\left|\Delta_R-\Delta_D\right|<\Omega<\Delta_R+\Delta_D$, where $\Delta_D\equiv 2\beta k_F$ and for definiteness we choose $\Delta_{R,D}>0$. The collective modes live below the lower bound of the continuum. The susceptibility is still a diagonal matrix so that the $1-2-3$ sectors are all decoupled. In fact, the non zero elements of $\hat\Pi^0$ are:
\bea\label{eq:D}
F_{1}(\Omega)&=&-\nu_{2D}\left[1+\frac{\Omega^2}{W_D^2}\right],\nonumber\\
F_{2}(\Omega)&=&-\nu_{2D}\frac{\Delta_R^2+\Delta_D^2}{2\Delta_R\Delta_D}
\left[1-\frac{1}{W_D^2}\left(-\Omega^2+\frac{(\Delta_R^2-\Delta_D^2)^2}{\Delta_R^2+\Delta_D^2}\right)\right],\nonumber\\
\Pi^0_{11}(\Omega)&=&F_1(\Omega)+F_2(\Omega),\nonumber\\
\Pi^0_{22}(\Omega)&=&F_1(\Omega)-F_2(\Omega),\nonumber\\
\Pi^0_{33}(\Omega)&=&2F_{1}(\Omega),
\eea
where $W_D^2=\sqrt{(-\Omega^2 + (\Delta_R+\Delta_D)^2)(-\Omega^2 + (\Delta_R-\Delta_D)^2)}$. The diagonal form of $\hat\Pi^0$ suggests that all the modes are linearly polarized. The frequencies can be found from the following three mode equations:
\bea\label{eq:D2}
&&1+UF_1(\Omega)=0,\\
&&1\pm\frac{u(\Delta_R\mp\Delta_D)^2}{4\Delta_R\Delta_D}
\left(1-\sqrt{\frac{(\Delta_R\pm\Delta_D)^2-\Omega^2}{(\Delta_R\mp\Delta_D)^2-\Omega^2}}\right)=0,
\eea
where the order of signs is to be maintained through the equation. Solving them, we get the collective mode frequencies:
\bea\label{eq:M12}
&&\Omega_i^2=(\Delta_R-\Delta_D)^2\left[1- \frac{uz_i}{2z_1} \right],~
i\in (1,2);\nonumber\\
&&z_1=\left[1+\frac{2\Delta_R\Delta_D}{u(\Delta_R-\Delta_D)^2}\right]^{-1},~z_2=1-\frac{2\Delta_R\Delta_D}{u(\Delta_R+\Delta_D)^2},\nonumber\\
&&\Omega_3^2=(\Delta_R-\Delta_D)^2\left[1-\frac{u^2}{1-2u}\frac{\sqrt{1+z_3 + z_3\frac{\Delta_R^2+\Delta_D^2}{2\Delta_R\Delta_D}}-(1+z_3)}{z_3}\right],\nonumber\\
&&z_3=\left(\frac{u}{1-u}\right)^2\frac{(\Delta_R-\Delta_D)^2}{2\Delta_R\Delta_D}.
\eea
Here $z_2\in(0,1)$. These solutions are plotted in Fig. \ref{fig:RDB}right as function of increasing DSOC. It is clear from the above equations that $\Omega_1$ and $\Omega_3$ graze the continuum up to the gap-closing point, where as $\Omega_2$ hits the continuum when $z_2=0$ before the gap-closing point. Here we have 3 collective modes on each side of this point. Further, there is complete symmetry in $\Delta_R\leftrightarrow \Delta_D$. This property does not hold when DSOC is replaced with $B$ as it is intimately connected to the appearance of $C_2$ preserving $\sin2\theta_k$ factor which as opposed to the rotational symmetry breaking $\sin\theta_k$ factor in the case of $\vec{B}$. As we increase $B$ from zero, we start with one/two/three collective modes (depending on the value of $\Delta_D/\Delta_R$),  we pass through the closing and reopening of the continuum gap as a function of $B$, and enter the regime where three is only one precessing mode(as shown in Fig. \ref{fig:RDB}left).

\section{Analysis of the eigenmode equations for the case of Rashba spin-orbit coupling and magnetic field present}
In this section, we analyze some properties of the eignemode equations for  the case when RSOC and magnetic field are present.
\subsection{Proving the non-existence of the $11$-sector collective mode for $B>B_c$}
The frequency of the collective mode in the $11$-sector (corresponding to oscillations of magnetization along the $x_1$ axis, i.e., along the static magnetic field) is determined from Eq.~(8a) of the MT, copied here for the reader's convenience:
$1+ \frac{U}{2}\Pi^0_{11}(\Omega)=0$. Here we prove that this equation has no
solution for $B>B_c$. Explicitly, this equation reads:
\beq\label{eq:app5}
\frac1u = \frac{(1-f)W^2}{4\tilde\Delta^2_Z},
\eeq
where
\beq
f\equiv\sqrt{1-4\Delta_R^2\tilde{\Delta}_Z^2/W^4}\label{def:f}
\eeq and
\beq
W^2\equiv  \Delta_R^2 + \tilde{\Delta}_Z^2-\Omega^2-i0^+\text{sgn}\Omega.
\label{def:W}
\eeq
Using the standard inequality of arithmetic and geometric means, we find that $f$ is always real and $<1$ if we restrict ourselves to the region below the continuum boundaries (i.e. $\Omega<|\tilde\Delta_Z-|\Delta_R||$). This implies that the right-hand side (RHS) of Eq.~(\ref{eq:app5}) is smaller than $\frac{W^2}{4\tilde\Delta^2_Z}$ which, can be immediately seen to be less than $\frac12$ for $\frac{|\Delta_R|}{\tilde\Delta_Z}<1$, i.e, for $B>B_c$. Therefore, we have $0<$RHS$<1/2$, while the left-hand side is larger than $1$ within the paramagnetic phase ($u<1$). Thus there is no solution of Eq.~(\ref{eq:app5}) for $B>B_c$.

\subsection{Collective modes in the $22$ and $33$ sectors}
We now analyze Eq.~(8b) in various limits to derive the results presented in Eqs.~(9-11) of the MT. Equation (8b) can be rewritten as
\bea\label{eq:app6}
\frac1u = X + \frac{1}{1-u} Y,
\eea
where
\bea
X&\equiv&\frac{\tilde\Delta^2_Z}{fW^2}-\frac14\left(\frac1f-1\right)\left(3-\frac{\Delta^2_R}{\tilde\Delta^2_Z}\right),\\
Y&\equiv&\frac{\Omega^2}{\tilde\Delta^2_Z}\left[\frac{\tilde\Delta^2_Z}{fW^2}-\frac14\left(\frac1f-1\right)\right].
\eea
To proceed further, we introduce the dimensionless quantities $w\equiv\Omega/\tilde\Delta_Z$ and $r\equiv |\Delta_R|/\tilde\Delta_Z$. Further, we look for solution of the form $w^2=(1-r)^2-\delta$ and solve for $0<\delta<(1-r)^2$.  In these notations, $W^2/\tilde\Delta_Z^2=2r+\delta$, $fW^2=\sqrt{\delta}\sqrt{4r+\delta}$. The quantities $X$ and $Y$ can be re-written as
\bea\label{eq:app7}
X&=&\frac{1}{ab}-\frac{3-r^2}{8}\frac{(a-b)^2}{ab},\nonumber\\
Y&=&\left[\frac{1}{ab}-\frac{1}{8}\frac{(a-b)^2}{ab}\right]\left[(1-r)^2-\delta\right],
\eea
where $a=\sqrt{\delta}$ and $b=\sqrt{4r+\delta}$. Its easy to see that in the limit $r\rightarrow 0$, $a\rightarrow b$ and $\delta = u(2-u)$. This makes $\Omega^2/\tilde\Delta_Z^2=(1-u)^2$ or $\Omega=\Delta_Z$, which is bare the Larmor frequency. In the opposite limit of $r\rightarrow \infty$, we find $\delta=r^2u$ or $r^2u/2$. These give $\Omega^2 = \Delta_R^2(1-u)$ or $\Delta^2_R(1-u/2)$, which are the two chiral-spin mode frequencies in the absence of the magnetic field.

Equation (9) in the MT denotes the strong field limit and is derived assuming $r\ll u<1$. We skip this derivation as it is a brute force expansion in $r^2$ which is lengthy but completely straightforward.

In the moderate field limit, where $r\approx 1$, we relabel $r=1-\e$ where $ 0<\e\ll 1$ and look for a solution in the interval $\delta\ll\e^2$. In this limit, the quantities $X$ and $Y$ reduce to
\bea\label{eq:app8}
X&=&\frac12+\frac{3\e^2}{4\sqrt{\delta}},\nonumber\\
Y&=&\frac{\e^2}{4\sqrt{\delta}},
\eea
This yields
\bea\label{eq:app9}
\delta&=&\frac{\e^4}{4}\frac{u^2(1-3u/4)^2}{(1-u)^2(1-u/2)^2},
\eea
which reproduces Eq.~(10) of MT.

In the weak-coupling case ($u\ll 1$), one can separate one more interval:
$u\ll r<1$. There, we find that $\delta=u^2(1-r)^4/r$. This makes the frequency $\Omega^2\approx\tilde\Delta_Z(1-r)\left[1-\frac{u^2}{2}\frac{(1-r)^2}{r}\right]$ which reproduces Eq.~(11) of  MT.

\section{Collective modes from the Quantum Kinetic Equation}\label{App:QKE}
The quantum kinetic equation for a Fermi liquid (FL) subject to a spatially uniform external filed and in the collisionless regime reads
\beq\label{eq:app10}
i\frac{\partial\delta\hat n_\bk}{\partial t} = [\delta\hat{\e}_\bk,\hat n_\bk],
\eeq
where
\beq
\delta\hat{\e}_\bk = \frac{\vec{s}_\bk\cdot\vec{\sigma}}{2} + \int' \text{Tr}'[\hat{F}_{\bk\bk'}\delta\hat n_\bk']
\eeq
is a variation of the quasiparticle energy, $F_{\bk\bk'}=F^a(\theta-\theta')\vec\sigma\cdot\vec\sigma'$ is the antisymmetric part of an
$SU(2)$-invariant Landau interaction function, $\theta$ and $\theta'$ are the angle subtended by $\vec k$ and $\vec k'$, correspondingly, and $\vec{s}_\bk$ parametrizes the spin-orbit and Zeeman terms of the Hamiltonian. For RSOC,
$\vec s_\bk=\Delta_R(\sin\theta,-\cos\theta,0)$; for purely Zeeman coupling,  $\vec s_\bk=(-\tilde\Delta_Z,0,0)$, etc. The electron distribution function can be written as
\beq
\hat n_\bk=\frac{\vec{s}_{\bk}\cdot\vec{\sigma}}{2}\frac{\partial n_0}{\partial\e} + \delta\hat n_\bk,
\eeq
where
$n_0$ is the equilibrium distribution function in the absence of both SOC and magnetic field and $\delta\hat n_\bk$ is the non-equilibrium part. The non-equilibrium part of the magnetization is given by
\beq
\label{magn}
\vec{M}=-\frac{g\mu_B}{2}\int_\bk \text{Tr}[\vec{\sigma}\delta\hat n_\bk].
\eeq

The non-equilibrium part of the distribution function can be expanded either over standard or rotated Pauli matrices \cite{Shekhter}. In the first way,
$\delta\hat n_\bk=\vec N(\theta)\cdot\vec\sigma\frac{\partial n_0}{\partial\e}$
such that $M_i=g\mu_B\nu_{2D}\int_{\theta}N_i(\theta)$, with $i\in(1,2,3)$ and $\int_\theta\equiv \int^{2\pi}_0 d\theta/(2\pi)$. The kinetic equation
reads\bea\label{eq:app11}
\dot{\vec N}(\theta) &=& -\vec N(\theta) \times\vec s_\theta -  \int_{\theta'}F^a(\theta-\theta')\vec N(\theta')\times\vec s_{\theta},
\eea
where $\vec s_\theta\equiv \vec s_{\vec k}$ at $k=k_F$. Note that the time dependence of $\vec N$ is not explicitly specified.
Equation (\ref{eq:app11}) can be solved  by decomposing $\vec N$ and $F^a$ into angular harmonics. Note that since $M_i$ is given by the zeroth harmonic of $N_i$.

As a demonstration, we solve Eq.~(\ref{eq:app11}) for the case of RSOC in the $s-$wave approximation for $F^a(\theta-\theta') = F^a_0$.  Equation (\ref{eq:app11})  is then simplified to
\beq\label{eq:apptemp}
\dot{\vec N}(\theta) = - \vec N(\theta) \times \vec s_\theta - F^a_0 \vec M \times \vec s_\theta.
\eeq
Note that $\vec s_\theta\cdot \dot{\vec N}(\theta) = 0$ suggesting that $\vec s_\theta\cdot {\vec N}(\theta)=$const, which can be set to zero. Integrating Eq.~(\ref{eq:apptemp}) over $\theta$ and noticing that $\int_\theta\vec s_\theta=0$ for RSOC, we get
\beq\label{eq:apptemp2}
\dot{\vec M} = - \int_\theta\vec N(\theta) \times \vec s_\theta.
\eeq
Differentiating Eq.~(\ref{eq:apptemp2}) over time again and using Eq.~(\ref{eq:apptemp}) for $\dot{\vec N}(\theta)$ with  $\vec s_\theta\cdot {\vec N}(\theta)=0$, we obtain \beq\label{eq:apptemp3}
\ddot{\vec M} = - \left(1+F_0^a\right)\Delta_R^2 \vec M + F_0^a \int_\theta \vec s_\theta \left(\vec s_\theta\cdot \vec M\right).
\eeq
This yields
\beq\label{eq:apptemp3.5}
\ddot{M}_{1,2} = - \left(1+\frac {F^a_0}{2}\right)\Delta_R^2 M_{1,2},~~~~ \ddot{M}_{3} = - \left(1+F_0^a\right)\Delta_R^2 M_{3},
\eeq
which coincides with $q=0$ limit of the hydrodynamic equations derived in Ref.~\onlinecite{Ali1}.

For the field only case, when $\vec{s}_\theta=(-\tilde\Delta_Z,0,0)$ is isotropic in the momentum space, we obtain the familiar Bloch equation by integrating Eq.~(\ref{eq:app11}) over the angle \cite{LL}
\beq\label{eq:apptemp4}
\dot{\vec M} =  (1+F_0^a)\tilde\Delta_Z\vec M \times \hat x_1 = g\mu_B \vec M\times\vec B.
\eeq
\subsection{Equivalence of the RPA and FL approaches in the $s$-wave approximation}
To analyze of the case RSOC in the presence of the magnetic field, it is beneficial to carry out a different   because equations for decomposition:
\bea\label{eq:app12}
\delta\hat n &=& N_i(\theta)\hat\tau_i\frac{\partial n_0}{\partial \e},\nonumber\\
\hat\tau_1 &=& \hat\sigma_3, ~~ \hat\tau_2 = \cos\theta\hat\sigma_1+\sin\theta\hat\sigma_2,~~\hat\tau_3 = \sin\theta\hat\sigma_1-\cos\theta\hat\sigma_2.
\eea
Expanding $N_i(\theta)$ into angular harmonics as
\beq N_i(\theta)=\sum_m N_1^{(m)}\cos m\theta + \bar N_1^{(m)}\sin m\theta
\eeq
and using Eq.~(\ref{magn}), we obtain for the magnetization components
\bea
M_1&=& g\mu_B\left(N^{(1)}_2 + \bar N_3^{(1)}\right),~~M_2=g\mu_B
\left(\bar N^{(1)}_2 -  N_3^{(1)}\right),~~M_3 =g\mu_B N_1^{(0)}.
\eea

The case of RSOC only can be solved exactly for an arbitrary form of the Landau interaction function in the spin channel, $F^a(\theta-\theta')$, because equations for harmonics of $\vec N$ decouple in this case \cite{Shekhter}. However, harmonics do not decouple in the presence of the field for an arbitrary Landau function, and thus an exact solution is not possible. To proceed further, we adopt the $s$-wave approximation, $F^a_0(\theta-\theta')=F^a_0$. In this case, the kinetic equation [Eq.~(\ref{eq:app10})] can be written as
\bea\label{eq:app13}
\dot N_{1}(\theta)+\tilde{\Delta}_{z}
\left[N_{2}(\theta)\sin\theta-N_{3}(\theta)\cos\theta\right]-\Delta_{R}N_{2}(\theta)&=&
\tilde F\left[\Delta_{R}(M_{2}\sin\theta+M_{1}\cos\theta)-\tilde{\Delta}_{z}M_{2}\right],\nonumber\\
\dot N_2(\theta)-\tilde{\Delta}_{z}N_{1}(\theta)\sin\theta+\Delta_{R}N_{1}(\theta)&=&
\tilde F(\tilde{\Delta}_{z}\sin\theta-\Delta_{R})M_{3},\\
\dot N_3+\tilde{\Delta}_{z}N_{1}(\theta)\cos\theta&=&
-\tilde F\tilde{\Delta}_{z}\cos\theta M_{3}.\nonumber
\eea
where $\tilde F\equiv F^a_0/g\mu_B\nu_{2D}$.
After a Fourier transform in time, we obtain for those harmonics of $N$ that are relevant for magnetization:
\bea\label{eq:app14}
N^{0}_{1}&=&\frac{\tilde F}{2W^2\tilde{\Delta}_z f}\left[i\Omega\left\{W^2(1-f)+2\tilde{\Delta}_z^2 \right\}M_{2}-2\tilde{\Delta}_z \left\{(\Omega^2+W^2f)M_{3}\right\}\right],\nonumber\\
N^{1}_{2}&=&\frac{\tilde F}{8\tilde{\Delta}_z^2 \Delta_R^2}\left[2W^2\Delta^{2}_{R}(f-1) - \left\{W^4(f-1)+2\tilde{\Delta}_z^2 \Delta_R^2 \right\} \right]M_{1},\nonumber\\
\bar{N}^{1}_{2}&=&\frac{\tilde F}{8\tilde{\Delta}_z^2 \Delta_R^2f}\left[\left\{2(1-f)(2\tilde{\Delta}_{z}^2\Delta_{R}^2-W^2(\tilde{\Delta}^{2}_{z}+\Delta^{2}_{R}))-
(2\tilde{\Delta}_z^2 \Delta_R^2f +W^4(f-1))\right\}M_{2}\right.\nonumber\\
&&~~~~~~~~~~\left.+2i\Omega(1-f)\left\{W^2\tilde{\Delta}_{z}-2\tilde{\Delta}_z \Delta_{R}^2\right\}M_{3}\right],\nonumber\\
N^{1}_{3}&=&\frac{\tilde F}{8\tilde{\Delta}_z^2 \Delta_R^2}\left[\left\{2W^2\tilde{\Delta}^{2}_{z}(1-f)+(2\tilde{\Delta}_z^2 \Delta_R^2+W^4(f-1))\right\}M_{2}+2iW^2\Omega\tilde{\Delta}_{z}(f-1)M_{3}\right],\nonumber\\
\bar N^{1}_{3}&=&\frac{\tilde F}{8\tilde{\Delta}_z^2 \Delta_R^2}\left[2\tilde{\Delta}_z^2 \Delta_R^2+W^4(f-1)\right]M_{1},
\eea
where $f$ and $W$ are given by Eqs.~(\ref{def:f}) and (\ref{def:W}), correspondingly.
Combining the left-hand sides of the equations above into components of $\vec M$, we obtain the eigenmode equation
\beq\label{eq:app15}
\left(
\begin{array}{ccc}
1-\frac {F^a_0}{2\nu_{2D}}\Pi^0_{11}(\Omega)&0&0\\
0&1-\frac{F^a_0}{2\nu_{2D}}\Pi^0_{22}(\Omega)&-\frac{F^a_0}{2\nu_{2D}}\Pi^0_{23}(\Omega)\\
0&-\frac{F^a_0}{2\nu_{2D}}\Pi^0_{32}(\Omega)&1-\frac{F^a_0}{2\nu_{2D}}\Pi^0_{33}(\Omega)
\end{array}
\right)
\left(
\begin{array}{c}
M_1\\M_2\\M_3
\end{array}
\right)=0,
\eeq
where $\Pi^0_{ij}(\Omega)$ are the same as in Eq.~(7) of MT. These are the same eigenmode equations as given by RPA, det$[1+\frac U2\Pi]=0$, upon
replacing $F_0^a\rightarrow -\nu_{2D}U$.

\section{Spin-orbit parameters in some 2D heterostructures}

This a short summary of the the SOC parameters measured in some quantum wells.
\begin{center}
\begin{tabular}{ |c|c|c|c|c|c|c|c| }
 \hline
 Material &n (10$^{11}$cm$^{-2}$)&$|g$-factor$|$& $\alpha$(meV{\AA})&$\Delta_R=\frac{2\alpha k_F}{g\mu_B}$&$\beta$(meV{\AA})  & $\Delta_D=\frac{2\beta k_F}{g\mu_B}$ & References \\
 \hline
 SiGe/Si/SiGe&5&2&0.055&33.7mT&-&-&\cite{Si_Exp}\\
 Mg$_x$Zn$_{1-x}$O/ZnO&2&1.9&0.7&0.28T&-&-&\cite{MgZn_exp}\\
 Cd$_{1-x}$Mn$_x$Te &3.5&1.6 + Mn&3.3&0.7-2T,&4.6&1-3T&\cite{CdMnTe}\\
 GaAs/AlGaAs&2.3&0.4&3&7T&0.5&1T&\cite{AlGaAs}\\
 GaAs/AlGaAs&5.8&0.3&1.5&7T&1.4&6T&\cite{AlGaAs2}\\
 InAs&21&8&65&20T&4&1.3T&\cite{InAlAs,g_InAsQW}\\
 InAs&12&8&60&14T&-&-&\cite{InAs2}\\
 In$_{1-x}$Ga$_{x}$As/In$_{1-y}$Al$_{y}$As&20&4&70&42T&-&-&\cite{InGaAs}\\
 \hline
\end{tabular}
\end{center}

The `$+$Mn' refers to a situation where the effective field experienced by the conduction electrons is enhanced by coupling to local magnetic spin, in this case, to those of Mn. This enhancement is $\sim 5$ in CdMnTe.

\end{widetext}

\begin{thebibliography}{5}
\bibitem{Stormer} H. L. Stormer, Z. Schlesinger, A. Chang, D. C. Tsui, A.C. Gossard, and W. Wiegmann, ``Energy Structure and Quantized Hall Effect of Two-Dimensional Holes'', Phys. Rev. Lett. \textbf{51}, 126 (1983).
\bibitem{Klitzing}D. Stein, K. v. Klitzing, and G. Weimann, ``Electron Spin Resonance on GaAs/Al$_x$Ga$_{1-x}$As Heterostructures'', Phys. Rev. Lett. \textbf{51}, 130 (1983).
\bibitem{ESRcurr}
Z. Wilamowski, H. Malissa, F. Schaeffler, and W. Jantsch, ``g-Factor Tuning and Manipulation of Spins by an Electric Current'',
Phys. Rev. Lett. \textbf{98}, 187203 (2007).
\bibitem{Shekhter}
A. Shekhter, M. Khodas, and A. M. Finkelstein, ``Chiral spin resonance and spin-Hall conductivity in the presence of the electron-electron interactions'',
Phys. Rev. B \textbf{71}, 165329 (2005).
\bibitem{CIESR}
Z. Wilamowski, W. Ungier, and W. Jantsch, ``Electron spin resonance in a two-dimensional electron gas induced by current or by electric field'', Phys. Rev. B \textbf{78}, 174423 (2008).
\bibitem{abs}
W. Ungier and W. Jantsch, ``Rashba fields in a two-dimensional electron gas at electromagnetic spin resonance'', Phys. Rev. B \textbf{88}, 115406 (2013).
\bibitem{R1}
E. I. Rashba, Sov. Phys. Solid State \textbf{2}, 1109 (1960).
\bibitem{bychkov}Yu. A. Bychkov and E. Rashba, ``Properties of a 2D electron gas with lifted spectral degeneracy", JETP Lett. \textbf{39}, 78 (1984); ``Oscillatory effects and the magnetic susceptibility of carriers in inversion layers'', J. Phys. C \textbf{17}, 6039 (1984).
\bibitem{Dresselhaus}G. Dresselhaus, ``Spin-Orbit Coupling Effects in Zinc Blende Structures'', Phys. Rev. \textbf{100}, 580 (1955).
\bibitem{Winkler} R. Winkler, ``Spin--Orbit Coupling Effects in Two-Dimensional Electron and Hole Systems'', (Springer Berlin /Heidelberg, 2003).
\bibitem{RE} E. I. Rashba and V. I. Sheka, ``Electric-Dipole Spin Resonance'', in {\em Landau Level Spectroscopy}, (North-Holland, 1991), edited by G. Landwehr and E. I. Rashba, p. 131.
\bibitem{R2}
E. I. Rashba and Al. L. Efros, ``Efficient electron spin manipulation in a quantum well by an in-plane electric field'', Appl. Phys. Lett. \textbf{83}, 5295 (2003).
\bibitem{R3}
Al. L. Efros and E. I. Rashba, ``Theory of electric dipole spin resonance in a parabolic quantum well'', Phys. Rev. B \textbf{73}, 165325 (2006).
\bibitem{LossEDSR}
M. Duckheim and D. Loss, ``Electric-dipole-induced spin resonance in disordered semiconductors'', Nature Phys. \textbf{2}, 195 (2006).
\bibitem{EDSR_AlAs}
M. Schulte, J. G. S. Lok, G. Denninger, and W. Dietsche, ``Electron Spin Resonance on a Two-Dimensional Electron Gas in a Single AlAs Quantum Well'',
Phys. Rev. Lett. \textbf{94}, 137601 (2005).
\bibitem{silin:1958}V. P. Silin, ``Oscillations of a Fermi liquid in a magnetic field'', Sov. Phys. JETP {\bf 6}, 945 (1958).
\bibitem{leggett:1970}A. J. Leggett, ``Spin diffusion and spin echoes in liquid 3He at low temperature'', J. Phys. C: Solid State Phys. {\bf 3}, 448 (1970).
\bibitem{Yafet} Y. Yafet, ``'$g$-Factors and Spin-Lattice Relaxation", in {\em Solid State Physics}, \textbf{14}, 92, edited by F. Seitz and D. Turnbull; Academic, New York (1963).
\bibitem{Starykh} R. Glenn, O. A. Starykh, and M. E. Raikh, ``Interplay of spin-orbit coupling and Zeeman splitting in the absorption lineshape of fermions in two dimensions'', \prb \textbf{86}, 024423 (2012).
\bibitem{Ali1}
A. Ashrafi and D. L. Maslov, ``Chiral Spin Waves in Fermi Liquids with Spin-Orbit Coupling'', Phys. Rev. Lett. \textbf{109}, 227201 (2012).
\bibitem{SM1}
S. Maiti, V. A. Zyuzin, and D. L. Maslov, ``Collective modes in two- and three-dimensional electron systems with Rashba spin-orbit coupling'',
Phys. Rev. B \textbf{91}, 035106 (2015).
\bibitem{DP} M. I. D'yakonov and V. I. Perel', ``Spin relaxation of conduction electrons in noncentrosymmetric
semiconductors'', Sov. Phys. Solid State \textbf{13}, 3023 (1972).
\bibitem{Ganichev} S. D. Ganichev and L. E. Golub, ``Interplay of Rashba/Dresselhaus spin splittings probed by photogalvanic spectroscopy--A review'', Phys. Stat. Sol. (b) \textbf{251}, 1801 (2014).
\bibitem{Meyerovich} A. E. Meyerovich and K. A. Musaelian, ``Zero-temperature attenuation and transverse spin dynamics in Fermi liquids. I. Generalized Landau theory'', J. Low. Temp. Phys. \textbf{89}, 781 (1992); ``Zero-temperature attenuation and transverse spin dynamics in Fermi liquids. II. Dilute Fermi system'', J. Low. Temp. Phys. \textbf{94}, 249 (1994); ``Zero-temperature attenuation and transverse spin dynamics in fermi liquids. III. Low spin polarizations'', J. Low. Temp. Phys. \textbf{94}, 789 (1994).
\bibitem{Ali2}
A. Ashrafi, E. I. Rashba, and D. L. Maslov, ``Theory of a chiral Fermi liquid: General formalism'',
Phys. Rev. B \textbf{88}, 075115 (2013).
\bibitem{comment_U} Accounting for a non $s-$wave interaction
yields no additional modes that couple to macroscopic fields but only
renormalizes the frequencies of the existing modes, see Sec.~\lq\lq Collective modes from Quantum Kinetic Equation\rq\rq\/ of the Supplementary Material.
\bibitem{supp_FL2} Supplementary material, Sec.~``Single Particle Hamiltonian: Eigen states and self energy correction''.
\bibitem{comment_alpha} In the $s$-wave approximation, $\alpha$ is not renormalized by the electron-electron interaction. In a more general treatment, $\alpha$ should be understood as a renormalized value of the Rashba parameter \cite{Shekhter,raikh:1999,saraga:2005}
\bibitem{raikh:1999} G.-H. Chen and M. E. Raikh, ``Exchange-induced enhancement of spin-orbit coupling in two-dimensional electronic systems'', \prb \textbf{60}, 4826 (1999).
\bibitem{saraga:2005}D. S. Saraga and D. Loss, ``Fermi liquid parameters in two dimensions with spin-orbit interaction'', \prb\textbf{72}, 195319 (2005).
\bibitem{supplement_3} See Sec.~``Collective Modes within the Random Phase Approximation'' of the Supplementary Material.
\bibitem{supplement_4} Supplementary material, Sec. ``Analysis of eigenmode equations for the case of Rashba spin orbit and magnetic field present''.
\bibitem{CdMnTe}
F. Baboux, F. Perez, C. A. Ullrich, I. D'Amico, G. Karczewski, and T. Wojtowicz, ``Coulomb-driven organization and enhancement of spin-orbit fields in collective spin excitations'', Phys. Rev. B \textbf{87}, 121303(R) (2013).
\bibitem{Sup} Supplementary material, Sec. ``Spin-orbit parameters in some 2D Heterostructures''.
\bibitem{ESRrange}
Y. Wiemann, J. Simmendinger, C. Clauss, L. Bogani, D. Bothner, D. Koelle, R. Kleiner, M. Dressel, and M. Scheffler, ``Observing electron spin resonance between 0.1 and 67 GHz at temperatures between 50 mK and 300 K using broadband metallic coplanar waveguides'', Appl. Phys. Lett. \textbf{106}, 193505 (2015).
\bibitem{VS1}
B. M. Norman, C. J. Trowbridge, J. Stephens, A. C. Gossard, D. D. Awschalom, and V. Sih, ``Mapping spin-orbit splitting in strained (In,Ga)As epilayers'',
Phys. Rev. B \textbf{82}, 081304(R) (2010).
\bibitem{VS2}
B. M. Norman, C. J. Trowbridge, D. D. Awschalom, and V. Sih, ``Current-Induced Spin Polarization in Anisotropic Spin-Orbit Fields'',
Phys. Rev. Lett. \textbf{112}, 056601 (2014).
\bibitem{VS4}
M. Luengo-Kovac, M. Macmahon, S. Huang, R. S. Goldman, and V. Sih, ``g-factor modification in a bulk InGaAs epilayer by an in-plane electric field'', Phys. Rev. B \textbf{91}, 201110(R) (2015).
Phys. Rev. Lett. \textbf{114}, 156803 (2015).
\bibitem{RB2}
E. Olshanetsky, J. D. Caldwell, M. Pilla, S.-C. Liu, C. R. Bowers, J. A. Simmons, and J. L. Reno, ``Temperature dependence and mechanism of electrically detected ESR at the $\nu=1$ filling factor of a two-dimensional electron system'',
Phys. Rev. B \textbf{67}, 165325 (2003).
\bibitem{RB1}
S. A. Vitkalov, C. R. Bowers, J. A. Simmons, and J. L. Reno, ``ESR Detection of optical dynamic nuclear polarization in GaAs/Al$_x$Ga$_{1-x}$As quantum wells at unity filling factor in the quantum Hall effect'',
Phys. Rev. B \textbf{61}, 5447, (2000).
\bibitem{EdESR}
C. F. O. Graeff, M. S. Brandt, M. Stutzmann, M. Holzmann, G. Abstreiter, and F. Schaeffler, ``Electrically detected magnetic resonance of two-dimensional electron gases in Si/SiGe heterostructures'',
Phys. Rev. B \textbf{59}, 13242 (1999).
\bibitem{EdESR2}
H. W. Jiang and Eli Yablonovitch, ``Gate-controlled electron spin resonance in GaAs/Al$_x$Ga$_{1-x}$As heterostructures'', Phys. Rev. B \textbf{64}, 041307(R) (2001).
\bibitem{LL} E. M. Lifshitz and L. P. Pitaevskii, {\em Statistical Physics}(Pergamon Press, New York, 1980).
\bibitem{Si_Exp}
Z. Wilamowski, W. Jantsch, H. Malissa, and U. R\"{o}ssler, ``Evidence and evaluation of the Bychkov-Rashba effect in SiGe/Si/SiGe quantum wells'', Phys. Rev. B \textbf{66}, 195315 (2002).
\bibitem{MgZn_exp}
Y. Kozuka, S. Teraoka, J. Falson, A. Oiwa, A. Tsukazaki, S. Tarucha, and M. Kawasaki, ``Rashba spin-orbit interaction in a Mg$_x$Zn$_{1−x}$O/ZnO two-dimensional electron gas studied by electrically detected electron spin resonance'', Phys. Rev. B \textbf{87}, 205411 (2013).
\bibitem{CdMnTe}
F. Baboux, F. Perez, C. A. Ullrich, I. D'Amico, G. Karczewski, and T. Wojtowicz, ``Coulomb-driven organization and enhancement of spin-orbit fields in collective spin excitations'', Phys. Rev. B \textbf{87}, 121303(R) (2013).
\bibitem{AlGaAs}F. Baboux, F. Perez, C. A. Ullrich, I. D’Amico, J. Gómez, and M. Bernard, ``Giant Collective Spin-Orbit Field in a Quantum Well: Fine Structure of Spin Plasmons'', Phys. Rev. Lett. \textbf{109}, 166401 (2012).
\bibitem{AlGaAs2} 
L. Meier, G. Salis, I. Shorubalko, E. Gini, S. Schoen, and K. Ensslin, ``Measurement of Rashba and Dresselhaus spin–orbit magnetic fields'', Nat. Phys, \textbf{3}, 650 (2007).
\bibitem{InAlAs}Y. H. Park, H.-J. Kim, J. Chang, S. H. Han, J. Eom, H.-J. Choi, and H. C. Koo, ``Separation of Rashba and Dresselhaus spin-orbit interactions using crystal direction dependent transport measurements'', App. Phys. Lett. \textbf{103}, 252407 (2013).
\bibitem{g_InAsQW}T. P. Smith III and F. F. Fang, ``g factor of electrons in an InAs quantum well'', Phys. Rev. B \textbf{35}, 7729 (1987).
\bibitem{InAs2}
J. P. Heida, B. J. van Wees, J. J. Kuipers, and T. M. Klapwijk and G. Borghs, 
``Spin-orbit interaction in a two-dimensional electron gas in a InAs/AlSb quantum well with gate-controlled electron density'', Phys. Rev. B \textbf{57}, 11911(1998). 
\bibitem{InGaAs}J. Nitta, T. Akazaki, H. Takayanagi, and T Enoki, ``Gate Control of Spin-Orbit Interaction in an Inverted In0.53Ga0.47As/In0.52Al0.48As Heterostructure'', Phys. Rev. Lett. \textbf{78}, 1335 (1997).


\end{thebibliography}
\end{document}